\documentclass[prl,aps,amssymb,twocolumn,floatfix]{revtex4}

\usepackage{graphicx}
\usepackage{dcolumn}
\usepackage{bm}

\begin{document}

\preprint{}

\title{Spin noise of electrons and holes in self-assembled quantum dots}

\author{S. A. Crooker$^{1*}$, J. Brandt$^2$, C. Sandfort$^2$, A. Greilich$^{2\dag}$, D. R. Yakovlev$^2$, D. Reuter$^3$, A. D. Wieck$^3$, M. Bayer$^2$}

\affiliation{$^1$National High Magnetic Field Laboratory, Los
Alamos, NM 87545  USA}

\affiliation{$^2$Experimentelle Physik II, Technische Universit\"{a}t Dortmund,
D-44221 Dortmund, Germany}

\affiliation{$^3$Angewandte Festk\"{o}rperphysik,
Ruhr-Universit\"{a}t Bochum, D-44780 Bochum, Germany}


\date{\today}
\begin{abstract}

We measure the frequency spectra of random spin fluctuations, or ``spin noise", in ensembles of (In,Ga)As/GaAs quantum dots (QDs) at low temperatures. We employ a spin noise spectrometer based on a sensitive optical Faraday rotation magnetometer that is coupled to a digitizer and field-programmable gate array, to measure and average noise spectra from 0-1 GHz continuously in real time with sub-nanoradian/$\sqrt{\textrm{Hz}}$ sensitivity.  Both electron \emph{and} hole spin fluctuations generate distinct noise peaks, whose shift and broadening with magnetic field directly reveal their \emph{g}-factors and dephasing rates within the ensemble. A large, energy-dependent anisotropy of the in-plane hole \emph{g}-factor is clearly exposed, reflecting systematic variations in the average QD confinement potential.

\end{abstract}

\maketitle
The dynamical properties of electron and hole spins in semiconductor quantum dots (QDs) are actively investigated for potential applications in spintronics and quantum information processing \cite{Loss}. Optical pump-probe studies have proven essential in this regard, directly revealing the \emph{g}-factors and coherence decays of spins in QD ensembles \cite{Gurudev, GreilichScience1, YugovaPRB} and in single QDs \cite{Berezovsky, AtatureNP, Press, Gerardot}. In principle, these important properties are also accessible via alternative measurement approaches based on ``spin noise'' spectroscopy, in which the intrinsic fluctuation spectra of the spins -- if measurable -- also reveal this dynamical information, in accord with the fluctuation-dissipation theorem \cite{Kubo}. In general, spin noise signals scale favorably as the number of measured spins $N$ decreases (falling only as $\sim$$\sqrt{N}$) \cite{CrookerNature, CrookerPRB, Jerschow}, suggesting their use as viable probes of few-spin systems. Indeed, single electron spin detection using ultra-sensitive cantilevers exploited spin noise \cite{Rugar}, and nuclear spin noise imaging \cite{Jerschow} of nm-scale specimens has been demonstrated \cite{DegenPNAS}. Using optical probes, electron spin noise spectra were recently measured in alkali vapors \cite{CrookerNature}, bulk GaAs \cite{OestreichPRL, CrookerPRB}, and in quantum wells \cite{MullerPRL}. However, spin noise spectroscopy of fully quantum-confined electrons and holes in QDs has not yet been achieved, due in part to the very small magnitude of their noise signatures which are typically much less than the background noise density from electronic and photon shot noise \cite{CrookerPRB}.  Considerable signal-averaging and efficient use of the available data stream are therefore essential aspects of these experiments.

Here we demonstrate spin noise spectroscopy to be a powerful probe of the dynamical properties of spins localized in semiconductor QD ensembles. Spin fluctuation spectra from 0-1~GHz are measured continuously in real time (no experimental dead time), using an optical Faraday rotation magnetometer coupled to a fast digital spectrum analyzer. The data exhibit sub-nanoradian/$\sqrt{\textrm{Hz}}$ sensitivity over the entire spectral range, revealing the tiny spin noise signatures from both quantum-confined electrons \emph{and} holes. The evolution of these noise peaks with magnetic field directly reveals their \emph{g}-factors and dephasing rates, and clearly exposes a marked, size-dependent anisotropy of the in-plane hole \emph{g}-factor $g_{h\perp}$ which -- in contrast to electron \emph{g}-factors -- has been predicted to be a sensitive probe of QD confinement potentials \cite{FlattePRL, Hawrylak, Kusrayev}.

Figure 1(a) shows the experiment. (In,Ga)As/GaAs QDs are grown by molecular beam epitaxy on (100) GaAs and then annealed \cite{YugovaPRB}. All structures contain 20 layers of QDs separated by 60~nm GaAs barriers, with $\sim$$10^{10}$ QDs/cm$^{2}$ per layer. The QDs are nominally undoped. The samples are mounted on the cold finger of an optical cryostat. Random spin fluctuations in the QDs, $\delta S_z(t)$, impart Faraday rotation fluctuations $\delta \theta_F(t)$ on a narrowband probe laser that is focused to a 3.5 $\mu$m spot on the sample. Balanced photodiodes detect $\delta \theta_F(t)$, and the amplified output voltage $\delta V(t)$ is sent to a 1~GHz real-time digital spectrum analyzer implemented in the parallel configurable logic of a field-programmable gate array (FPGA).

This FPGA-based approach to spectral analysis and averaging maximizes the data collection efficiency and therefore the noise sensitivity. In comparison, traditional spectrum analyzers employing ``swept" local oscillators ignore most of the available signal -- e.g., measuring a 0-1~GHz spectrum with 1~MHz resolution effectively discards $\sim$99.9\% of the relevant data stream at any given time. Improved throughput is achieved by digitizing and averaging fast-Fourier transforms (FFTs) of the data stream so that all frequencies are averaged simultaneously. However, the finite interface bandwidth between typical plug-in digitizers and host computers -- and the substantial computational demands on the host processor(s) -- can impose significant limitations \cite{limitations}.  Using standard digitizers to realize 1~GHz bandwidth, typically $<$5\% of the available signal is actually used, with a corresponding reduction in experimental sensitivity.

\begin{figure}[tbp]
\includegraphics[width=.46\textwidth]{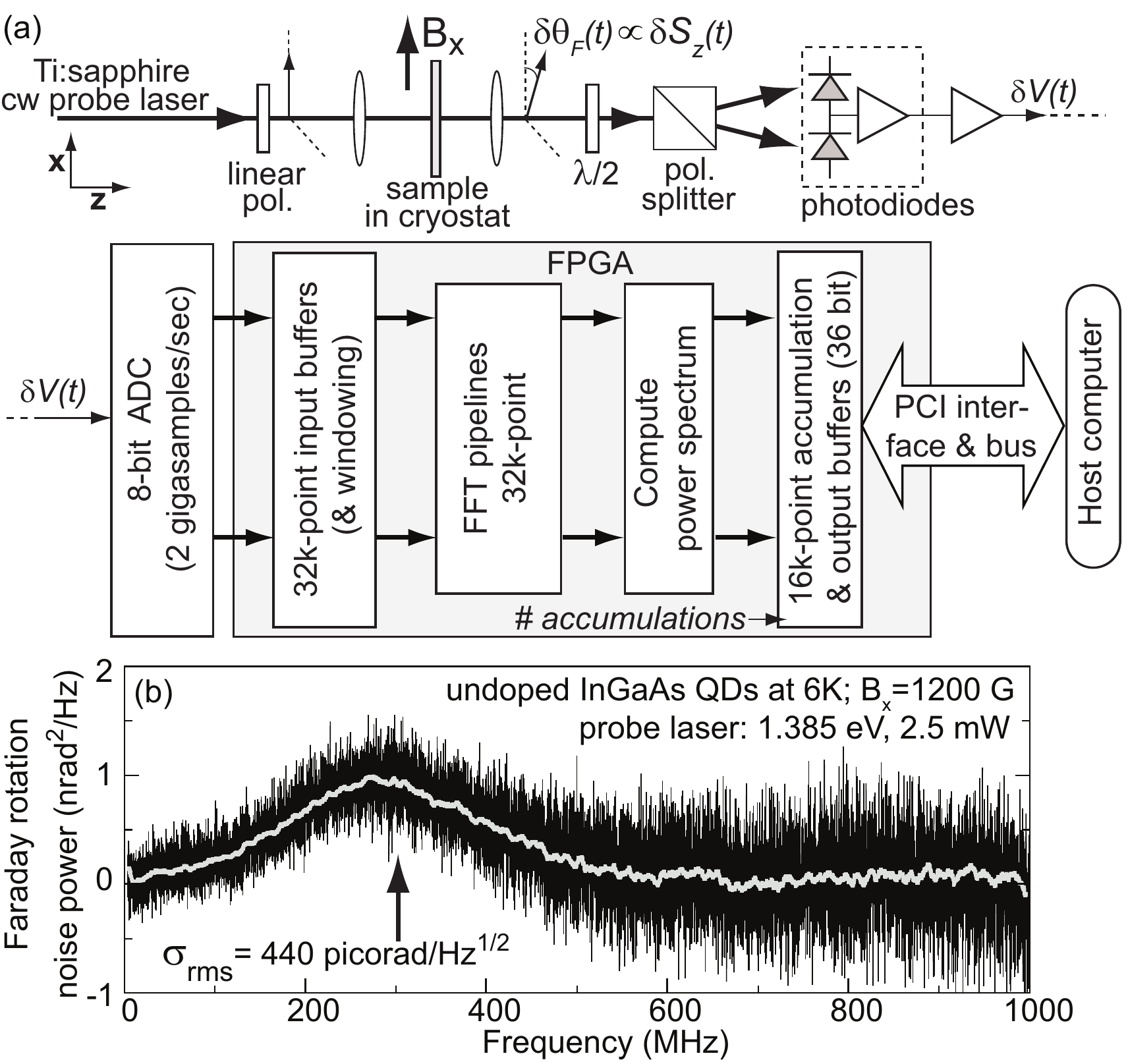}
\caption{(a) Experimental schematic. Spin fluctuations $\delta S_z(t)$ impart Faraday rotation fluctuations $\delta \theta_F(t)$ on the probe laser. $\delta V(t)$ is digitized at 2 GS/s, and its power spectrum from 0-1 GHz is computed and averaged continuously on a FPGA. (b) A raw noise power spectrum showing hole spin noise in (In,Ga)As QDs after 600~s averaging (1.2 TB of processed data; corrected for frequency-dependent detector gain). The rms noise floor at 300 MHz is 440 picorad/$\sqrt{\textrm{Hz}}$, which can be further reduced by smoothing (grey line).} \label{fig1}
\end{figure}

We avoid these limitations by using digitizers incorporating on-board FPGA processors to continuously compute and average power spectra from a 2 gigasample/s data stream in real time.  Recently developed for applications in radioastronomy and atmospheric physics \cite{Acqiris}, they typically serve as powerful digital back-ends to terahertz and millimeter-wave telescopes \cite{Klein}. However, the very weak broadband spectral signatures of spin noise in QDs impose a  similar need for fast, efficient computation and averaging of power spectra. Details of the spectrometer have appeared elsewhere \cite{Klein}, but are summarized in Fig. 1(a): $\delta V(t)$ is digitized at 2 GHz, and the FPGA continuously executes parallel 32k-point FFT pipelines and accumulates (averages) the computed power spectra on-board. Every second ($\sim$61000 averages), a single 16k-point power spectrum spanning 0-1~GHz is read by the host computer, and further averaging occurs in software.  Acquisitions of 1-100 minutes are typical, with 1 terabyte of data processed every 8.3 minutes.

To study spin noise, we apply in-plane magnetic fields $B_x$ to the QDs, forcing all spin fluctuations $\delta S_z$ to precess ($B_x \parallel [110]$ unless otherwise mentioned). This shifts the peak of the spin noise from zero to the relevant Larmor frequency $\omega_L = g \mu_B B_x/\hbar$. Also, we interleave (and subtract) spectra at $B_x$ with ``background" spectra acquired at a different field that shifts the spin noise out of the spectral range of interest, thus leaving behind only the noise due to fluctuating spins \cite{CrookerPRB, OestreichPRL}. Figure 1(b) shows one such Faraday rotation noise spectrum after 600~s of averaging. The noise floor of the raw data (black) is $<$1 nanoradian/$\sqrt{\textrm{Hz}}$ over the entire 1~GHz spectrum, and can be reduced further by smoothing over frequency bins (at the expense of spectral resolution; grey trace). The broad peak at $\sim$290 MHz is due to the spin noise of \emph{holes} in these QDs, as discussed next.

\begin{figure}[tbp]
\includegraphics[width=.45\textwidth]{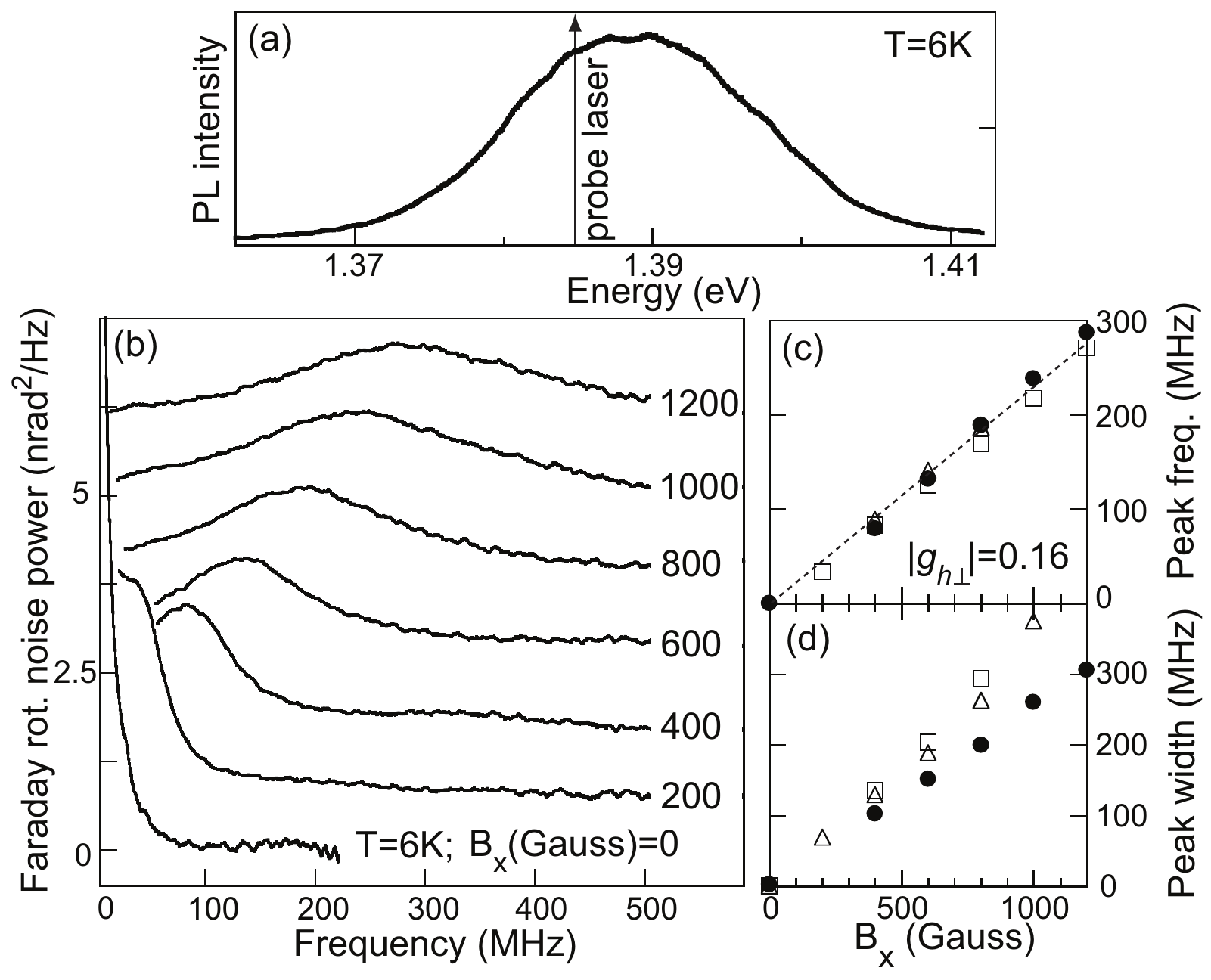}
\caption{(a) PL spectra of a QD ensemble (1.58 eV excitation). (b) Faraday rotation noise power spectra versus $B_x$ ($\parallel$ [110]), using 2.5 mW, 1.385 eV probe laser. (c,d) The hole spin noise peak frequency and width.} \label{fig2}
\end{figure}

Figure 2(a) shows photoluminescence (PL) spectra from a QD ensemble. While individual QDs can exhibit narrow linewidths $<$10 $\mu$eV \cite{AtatureNP}, the ensemble PL is broadened to $\sim$15 meV by the inhomogeneous QD size distribution. We tune the probe laser to within this distribution of QD ground state transitions, as depicted.  Figure 2(b) shows noise spectra from these QDs, that clearly reveal the presence of spin fluctuations: a noise peak is observed, which shifts and broadens with increasing $B_x$. The evolution of the peak's center frequency and width are shown in Figs. 2(c,d). Repeated studies on different sample regions (symbols) show that the noise peak shifts linearly with $B_x$ at $\sim$220 kHz/G (dashed line), indicating that these fluctuating spins precess with an in-plane $g$-factor of $|g_{\perp}|$$\simeq$0.16. This value is considerably less than the known in-plane $g$-factor of electrons in very similar (In,Ga)As QDs ($|g_{e\perp}|$=0.54-0.57) which has been accurately established by pump-probe studies \cite{GreilichScience1,YugovaPRB}. Rather, it corresponds to the in-plane \emph{g}-factor of \emph{holes} in these QDs, measured in Ref. \cite{YugovaPRB} to be $|g_{h\perp}| = 0.15 \pm 0.05$.

The full width of the hole spin noise peak, $\Delta \omega_h$, which indicates dephasing rate, increases linearly with $B_x$ [Fig. 2(d)]. This directly reveals that an inhomogeneous distribution of hole $g$-factors, $\Delta g_h$, is probed in this measurement, leading to a spread of hole precession frequencies and, consequently, to rapid ensemble dephasing of hole spins in $B_x$. The ensemble spin dephasing time $T_{2,h}^*$ is given by $2/T_{2,h}^*=\Delta \omega_h = \Delta g_h \mu_B B_x/\hbar$, from which we estimate that $\Delta g_h$$\sim$0.2, such that $T_{2,h}^*$$\sim$1~ns at $B_x$=1000~G.

These data represent not only the first demonstration of spin noise spectroscopy of carriers confined in zero-dimensional QDs, but also constitute the first observation of hole spin noise in any semiconductor system. Direct measurements of $g_{h}$ in QDs are particularly important because, unlike electron \emph{g}-factors, the magnitude and anisotropy of $g_h$ are predicted to be especially sensitive to the specific details of the QD confinement potential \cite{FlattePRL, Hawrylak, Kusrayev}. The utility of these noise techniques are further exemplified in the following studies, which address this prediction explicitly: Firstly, hole noise  reveals a substantial in-plane anisotropy of $g_{h\perp}$ in these QDs.  Figure 3(a) shows hole spin noise spectra (acquired at 1.386~eV) shifting to lower frequencies as the [110] sample axis is rotated by $\theta$=0-90$^\circ$ from $B_x$. Linear fits versus $B_x$ indicate that $|g_{h,110}|$=0.165$\pm$.005 and $|g_{h,1\bar{1}0}|$=0.065$\pm$.005, revealing a strong in-plane anisotropy $P=(g_{110}-g_{1\bar{1}0})/(g_{110}+g_{1\bar{1}0})$$\simeq$43\%, which is over an order of magnitude larger than the 2.7\% anisotropy of \emph{electron} \emph{g}-factors previously measured in similar dots \cite{YugovaPRB}. Figure 3(b) shows this hole anisotropy explicitly, where the line is given by $g^2_{h\perp}=g_{110}^2 \textrm{cos}^2\theta + g_{1\bar{1}0}^2 \textrm{sin}^2\theta$. Thus we find that, indeed as postulated \cite{FlattePRL, Hawrylak}, hole \emph{g}-factors are considerably more sensitive than electron \emph{g}-factors to in-plane distortions of QD confinement potentials, which can arise from, \emph{e.g.}, shape or strain anisotropy.

\begin{figure}[tbp]
\includegraphics[width=.48\textwidth]{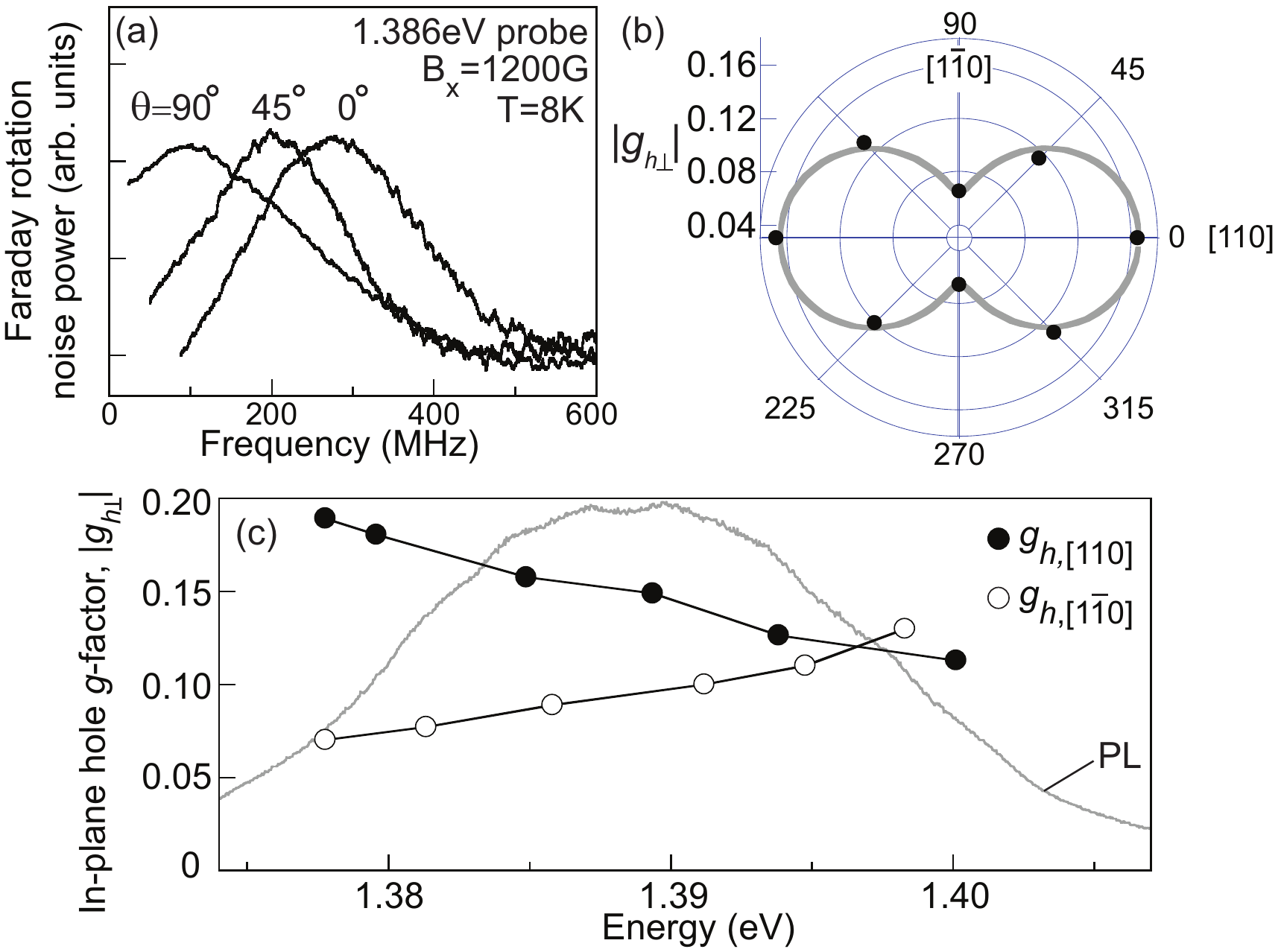}
\caption{(a) Hole spin noise with [110] axis rotated 0$^\circ$, 45$^\circ$, and 90$^\circ$ from $B_x$. (b) In-plane anisotropy of $g_{h\perp}$ at 1.386~eV. (c) Energy-dependent anisotropy of $g_{h\perp}$.} \label{fig3}
\end{figure}

Secondly, these noise studies unexpectedly reveal that this anisotropy itself varies markedly and systematically within the inhomogeneous distribution of QDs.  Figure 3(c) shows that $|g_{h,110}|$ and $|g_{h,1\bar{1}0}|$ exhibit \emph{opposite} trends as a function of increasing probe energy, such that the anisotropy of $g_{h\perp}$ in these QDs varies from $\sim$+50\% to -10\%, inverting sign at 1.397~eV. This surprising behavior strongly indicates that lateral shape or strain anisotropy of these self-assembled (In,Ga)As QDs, which depend sensitively on growth and annealing conditions, nonetheless varies in a systematic way as the QD confinement energy increases within a given sample.

Interestingly, corresponding signatures of \emph{electron} spin noise in these QDs are found to be very faint, being barely visible in the raw data. However, when the QDs are \emph{also} weakly illuminated with above-barrier light (at 1.58 eV), electron spin noise becomes quite apparent. Under these conditions (see Fig. 4), a clear peak shifts linearly with $B_x$ at $\sim$760 kHz/G, giving an in-plane $|g_\perp|$=0.54$\pm$.02. This value lies within the range of transverse electron \emph{g}-factors $|g_{e\perp}|$=0.54-0.57 measured in very similar (In,Ga)As QDs using standard pump-probe techniques \cite{GreilichScience1,YugovaPRB}. In contrast to the hole spin noise, Fig. 4(a) shows that the width of the electron spin noise peak remains relatively large (250-350 MHz) as $B_x$$\rightarrow$0, indicating a low-field electron spin dephasing time $\tau_e$ on the order of 1~ns. This ensemble $\tau_e$ is governed by the precession of each electron about the randomly-oriented nuclear hyperfine field $B_N$ in each QD \cite{Merkulov}, estimated here from $\tau_e$ to be $B_N = \hbar/\tau_e g_e \mu_B \sim$20 mT.

\begin{figure}[tbp]
\includegraphics[width=.45\textwidth]{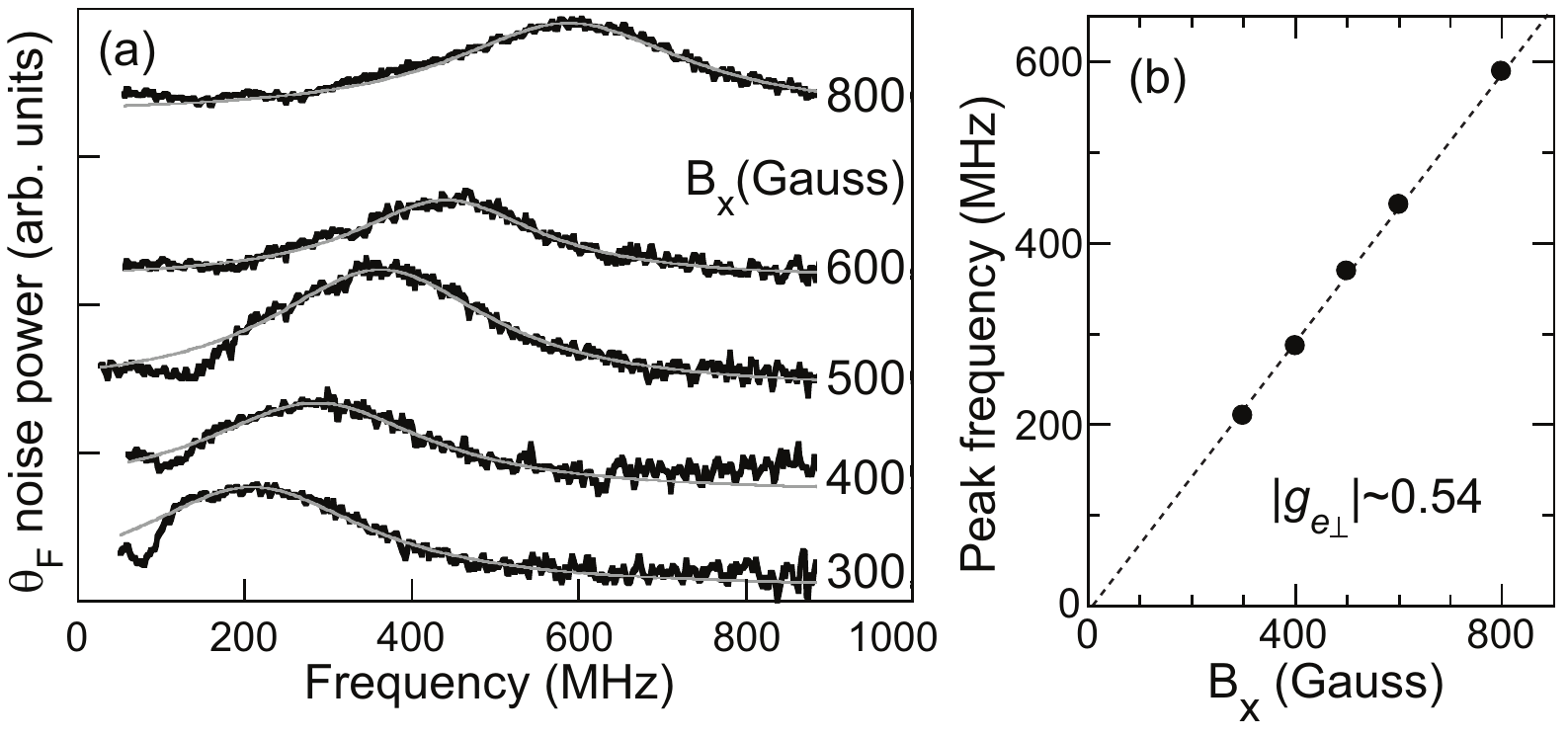}
\caption{(a) Electron spin noise, using 3 mW, 1.377 eV probe laser and additional 1.58 eV illumination. Grey lines are Lorentzian fits; $B_x$$\parallel$[110]. (b) Peak frequency versus $B_x$.} \label{fig4}
\end{figure}

Finally, for both electrons and holes, the magnitude of the spin noise signals are observed to be largest at the lowest temperatures (4-8~K), falling thereafter with increasing temperature to below our resolution limit for $T>30$~K. We also note that no spin noise is observed when the probe laser is tuned outside the PL band.

We turn now to the origin of the spin noise signals observed in these QD ensembles. Electron and hole spin fluctuations along $\hat{z}$ impart Faraday rotation on the probe laser via the usual spin-dependent optical selection rules for right- and left-circularly polarized light ($\sigma^\pm$), primarily through the difference between the QD's indices of refraction ($\theta_F \propto n^+ - n^-$). Both neutral excitons as well as single (resident) electrons and holes can give rise to spin noise in QDs. Consider the first case, wherein neutral excitons are directly excited by the probe laser itself: In these studies, the probe laser cannot be considered merely a passive observer of spin fluctuations (in contrast to noise studies in alkali vapors \cite{CrookerNature} and to a lesser extent in bulk $n$-GaAs \cite{CrookerPRB}, where the probe was detuned from absorption features). Rather, because the probe is tuned directly within the distribution of ground-state transitions, superpositions of spin-up and spin-down neutral excitons are directly pumped into the subset of QDs that are resonant with the probe laser. Spin fluctuations of the constituent holes and electrons can then generate Faraday rotation noise through the $\sigma^\pm$ selection rules for neutral excitons.  This scenario, however, cannot explain the marked difference in the magnitudes of the observed hole and electron spin noise.

Alternatively, spin fluctuations of single holes and electrons in the (nominally undoped) QDs can also generate Faraday rotation noise, via the $\sigma^\pm$ selection rules for positively- or negatively-charged trion transitions \cite{GreilichScience1, YugovaPRB, Berezovsky, AtatureNP, Press}. Because of the dispersive nature of the indices $n^\pm$, the probe laser will be most sensitive to spin fluctuations in those QDs having trion transitions near the probe laser energy (ideally, half an absorption linewidth away).  Although the QDs are not intentionally doped, a ubiquitous \emph{p}-type background doping from carbon impurities exists at a level of $\sim 2\times 10^{14}$/cm$^3$, which corresponds to $\sim$$10^9$ holes per QD layer, or 10\% of the QD density. Thus, the QDs are likely very weakly \emph{p}-type, which explains both the presence of hole spin noise and the comparative absence of electron spin noise in our data.  From these values, we estimate that $<$100 holes reside within, and are quasi-resonant with, the focused probe laser. Moreover, previous studies of two-dimensional hole gases in GaAs and in CdTe-based quantum wells \cite{Zhukov,Syperek} have shown that above-barrier illumination can deplete hole densities and even invert the majority carriers from holes to electrons. Our data is also consistent with these findings -- electron spin noise signals increase markedly with additional above-barrier illumination as described above, while hole noise signals decrease (not shown). 

Finally, the presence of isolated holes in the QDs can explain the remarkably narrow width of the hole spin noise as $B_x \rightarrow 0$ (Fig. 2), which indicates very long hole spin relaxation times $\tau_h$ of order tens of nanoseconds, greatly exceeding the 500~ps lifetime of neutral excitons. These noise data therefore provide compelling evidence in support of recent measurements of long intrinsic spin relaxation times of quantum-confined holes \cite{Eble, Gerardot}.

In summary, these spin noise data conclusively demonstrate that the dynamics of electron and hole spins in self-assembled QDs can indeed be measured by their fluctuation properties alone, using Faraday rotation methods and an efficient FPGA-based approach to broadband spin noise spectroscopy.  The noise data directly reveal a substantial and size-dependent anisotropy of the in-plane hole \emph{g}-factor $g_{h\perp}$, confirming recent theoretical expectations \cite{FlattePRL, Hawrylak} that \emph{g}-factor anisotropies of quantum-confined holes, as opposed to electrons, are considerably more sensitive probes of the specific details of confinement potentials in QDs. The observed electron and hole spin noise signatures are rather small in the present studies (noise power densities of order 1 nrad$^2$/Hz), but nonetheless provide a clear alternative route for measuring \emph{g}-factors, dephasing times, and symmetries of electron and hole spins confined in semiconductor QDs.

We thank Darryl Smith for helpful discussions, and acknowledge support from the Los Alamos LDRD program, the NHMFL, and the Deutsche Forschungsgemeinschaft.

$\dag$Current address: Naval Research Laboratory, Washington D.C. 20375, USA


\begin{references}
\bibitem{limitations}Standard 32-bit, 33 MHz PCI buses can transfer $\sim$$10^8$ 8-bit samples/s, limiting real-time bandwidths to $\sim$50 MHz (25 MHz for 16-bit samples) -- even if the host computer can compute and average FFTs at this rate. Larger desired bandwidths necessitate some ``dead time", during which the signal is ignored while data are transferred.
\end{references}
\end{document}